\documentclass[a4paper,twoside]{article}

\usepackage{comment,etoolbox}
\usepackage[ruled,vlined]{algorithm2e}
\usepackage{thm-restate}
\usepackage{listings}
\usepackage{epsfig}
\usepackage{stmaryrd}
\usepackage{calc}
\usepackage{amssymb}
\usepackage{amstext}
\usepackage{amsmath,leftidx}
\usepackage{graphicx}
\usepackage{thmtools,amsxtra,mathtools,nccmath,cleveref}
\usepackage{multirow,multicol,wrapfig,subfig}
\usepackage{pslatex}
\usepackage{booktabs,tabularray}
\usepackage{comment}
\usepackage{xargs}
\usepackage[colorinlistoftodos,prependcaption,textsize=footnotesize]{todonotes}
\usepackage{listings}
\usepackage[normalem]{ulem}%\sout \xout
\usepackage{soul}%\hl

\usepackage[bottom]{footmisc}
\usepackage{SCITEPRESS}     % Please add other packages that you may need BEFORE the SCITEPRESS.sty package.

\declaretheorem[name=Definition]{dft}

\declaretheorem[name=Lemma]{lmm}
 
\declaretheorem[name=Corollary]{cor}
\declaretheorem[name=Assumption]{ass}

\crefname{dft}{definition}{definitions}
\crefname{prp}{proposition}{propositions}
\crefname{con}{conjecture}{conjectures}
\crefname{axm}{axiom}{axioms}
\crefname{rem}{remark}{remarks}
\crefname{thm}{theorem}{theorems}
\crefname{lmm}{lemma}{lemmas}
\crefname{ass}{assumption}{assumptions}
\crefname{cor}{corollary}{corollaries}
\crefname{lst}{listing}{listings}
\crefname{exm}{example}{examples}

\usepackage[american]{babel}
\usepackage{csquotes}
\usepackage[
    backend=biber,
    style=apa,
  ]{biblatex}
  
\AtEveryBibitem{%
  \clearfield{issn} % Remove issn
  \clearfield{doi} % Remove doi

  \ifentrytype{online}{}{% Remove url except for @online
    \clearfield{url}
  }
}

\begin{document}

\title{Improving unlinkability in C-ITS: a methodology for optimal obfuscation}

\author{\authorname{Yevhen Zolotavkin\sup{1}\orcidAuthor{0000-0002-1875-122X}, Yurii Baryshev\sup{2}\orcidAuthor{0000-0001-8324-8869}, Vitalii Lukichov\sup{2}\orcidAuthor{0000-0002-3423-5436}, Jannik M\"{a}hn\sup{1}\orcidAuthor{0000-0003-0870-7193} and Stefan K\"{o}psell\sup{1}\orcidAuthor{0000-0002-0466-562X}}
\affiliation{\sup{1}Barkhausen Institut gGmbH, W\"{u}rzburger Straße 46, Dresden, Germany}
\affiliation{\sup{2}Department of Information Protection, Vinnytsia National Technical University, Khmelnytske Shosse 95, Vinnytsia, Ukraine}
\email{\{yevhen.zolotavkin, jannik.maehn, stefan.koepsell\}@barkhauseninstitut.org, \{yuriy.baryshev, lukichov.vitalyi\}@vntu.edu.ua}
}

\keywords{privacy, V2X, unlinkability, hidden Markov model, cybersecurity, entropy, obfuscation}

\abstract{In this paper, we develop a new methodology to provide high assurance about privacy in Cooperative Intelligent Transport Systems (C-ITS). Our focus lies on vehicle-to-everything (V2X) communications enabled by Cooperative Awareness Basic Service. Our research motivation is developed based on the analysis of unlinkability provision methods indicating a gap. To address this, we propose a Hidden Markov Model (HMM) to express unlinkability for the situation where two cars are communicating with a Roadside Unit (RSU) using Cooperative Awareness Messages (CAMs). Our HMM has labeled states specifying distinct origins of the CAMs observable by a passive attacker. We then demonstrate that a high assurance about the degree of uncertainty (e.g., entropy) about labeled states can be obtained for the attacker under the assumption that he knows actual positions of the vehicles (e.g., hidden states in HMM). We further demonstrate how unlinkability can be increased in C-ITS: we propose a joint probability distribution that both drivers must use to obfuscate their actual data jointly. This obfuscated data is then encapsulated in their CAMs. Finally, our findings are incorporated into an obfuscation algorithm whose complexity is linear in the number of discrete time steps in HMM.}

\onecolumn \maketitle \normalsize \setcounter{footnote}{0} \vfill

\section{\uppercase{Introduction}}\label{sec:introduction}

Due to the intense development of \textit{transport systems} over the recent decades, different modes of cooperative intelligence have been incorporated into their functionalities. \textit{Intelligent transport systems} (ITS) are transport systems in which advanced information, communication, sensor and control technologies, including the Internet, are applied to increase safety, sustainability, efficiency, and comfort. \textit{Cooperative Intelligent Transport Systems} (C-ITS) are a group of ITS technologies where service provision is enabled by, or enhanced by, the usage of `live', present situation related, dynamic data/information from other entities of similar functionality, and/or between different elements of the transport network, including vehicles and infrastructure (\cite{iso/tc204CooperativeITSPart2015}).

Technology allowing a vehicle to exchange additional information with infrastructure, other vehicles and other stakeholders in the context of C-ITS is called \textit{vehicle-to-everything} (V2X). Multiple advances in modern C-ITS applications, such as collaborative forward collision warning and emergency electronic brake lights, are impossible without V2X. These advances, however, come at a cost: C-ITS applications rely on vehicles broadcasting signals to indicate their location, signals which are intended to be received and processed by a range of other devices. For example, vehicles may cooperatively broadcast (with the frequency of 1-10 Hz) geo-spatial information to nearby peers using short Cooperative Awareness Messages (CAMs). Hence, V2X raises essential \textit{privacy questions}: {\bf \textit{i)}} to what degree can specific vehicles be located and tracked based on such information? {\bf \textit{ii)}} what are the techniques able to improve privacy of V2X? To answer these questions, we use the concept of \textit{unlinkability} to reason about privacy.

\subsection{Research motivation}

Even though the problems of privacy in C-ITS were acknowledged in several relevant documents (having normative and informative character), satisfactory answers have yet to be provided to the privacy questions mentioned above. For example, the document (\cite{iso/tc204ITSStationManagement2018}) recognizes the importance of unlinking private data from traceable address elements and identifiers in wireless messages sent by an ITS station unit (ITS-SU). However, relevant considerations in this document do not go beyond suggesting that ``such unlinking can be done by means of pseudonyms'': the sufficiency of these and many similar suggestions remain unaddressed. In contrast, limitations of pseudonym changes in CAMs have been recognized by academic authors (\cite{wiedersheimPrivacyIntervehicularNetworks2010,karimemaraBeaconbasedVehicleTracking2013,escherHowWellCan2021a}). In particular, vehicle tracking becomes possible due to CAM content being signed but \textit{not encrypted}: this is demanded by the relevant standards in C-ITS(\cite{itswg1IntelligentTransportSystems2019}). This is because full encryption of CAMs may impede C-ITS functionalities that are critical for safety. Nonetheless, recognition of privacy-affecting issues in C-ITS has yet to result in a solution where the degree of privacy is correctly measured, and privacy limitations are eliminated. In this paper, a methodology to address these challenges is developed: we provide an assurance for the procedure estimating the lower bound of unlinkability for CAMs and an algorithm maximizing this criterion under the overall constraint of location precision degradation.

\subsection{Research principles}

To address V2X privacy questions, it is imperative to obtain results for which confidence is high. The methodology developed in this paper rests on the principles of robust optimization (RO) (\cite{gorissenPracticalGuideRobust2015,sniedovichWaldMightyMaximin2016}). The unsuccessfulness of the previous attempts to answer the mentioned above `V2X privacy questions' is mainly caused by inopportune privacy assumptions for C-ITS and V2X scenarios in particular. For example, a combination of assumptions that an attacker only observe CAMs containing obfuscated (distorted) geo-position measurements and does not have a precise physical model for car movements may require reasoning involving the best-known technique to estimate such a model (\cite{blackmanMultipletargetTrackingRadar1986}). Because of the computational complexities associated with these estimations, researchers often rely on heuristic and poorly justifiable steps: this is unacceptable if high level of privacy assurance is demanded (\cite{blackmanMultipleHypothesisTracking2004,wiedersheimPrivacyIntervehicularNetworks2010}). To avoid this situation, we make stronger assumptions about the information known to the attacker, which allows us to estimate a lower bound of unlinkability in C-ITS: this estimate has high confidence. This attitude is reflected in the following \textit{principles} shaping our methodology:

\begin{itemize}
\item The methodology for privacy risk assessment should not underestimate the risks: it should provide clear and quantifiable assessments of \textit{how easy} data belonging to the same entity can be extracted throughout multiple V2X communication sessions involving more than one user. Therefore, the methodology should set \textit{justifiable bounds} for assessments of such quantity;

\item The methodology should propose \textit{optimal modifications} of the original user data. The notion of `optimality' depends on the kind of modification. Reversible modifications (such as encryption) should be provided with the assurances of computational infeasibility of such reversions by illegitimate parties (e.g., the level of such assurance is maximal). For irreversible modifications (such as noising), the loss of data quality should be quantified: for each such quantity, there should be an assurance that no higher degree of unlinkability can be achieved (e.g., the proposed noising method is optimal).  
\end{itemize}

\subsection{Our contribution}

The unique contribution is  due to combination of the study objective (guided by the criterion of unlinkability), robust assumptions, and the optimal obfuscation technique developed in the paper.

\begin{itemize}

\item First, the aim of this study is to protect C-ITS from the \textit{threat of linking}: this is in contrast with the numerous obfuscation approaches which aim at impeding inferencing about the actual location of ITS-SU (\cite{andresGeoindistinguishabilityDifferentialPrivacy2013,bordenabeOptimalGeoIndistinguishableMechanisms2014});
\item Second, to obtain \textit{high confidence} in the measured unlinkability, we assume that an attacker has complete knowledge about the system design, obfuscation algorithms, quality degradation (distortion) constraints, has access to CAMs, stored states vehicles' geo-positions, and the true states characterizing the geo-positions of the vehicles at any moment in time;
\item Third, we develop an \textit{optimal obfuscation} algorithm: for a given distortion constraint, it provides the highest level of uncertainty for the attacker trying to link obfuscated CAMs with their sources.   

\end{itemize}

The rest of this paper is structured as follows. In \cref{sec:prelim}, we set the grounds for the study and provide basic definitions. It is followed by \cref{sec:m_mod}, where we start with a description of a generic information system. Further specifics of C-ITS are then reflected using additional sets and relations: as a result, we obtain Hidden Markov Model (HMM), used to study unlinkability. In \cref{sec:mod_prop}, we formalize assumptions, define unlinkability through entropy and optimize joint obfuscation producing observable states in HMM. Next, \cref{sec:obf_alg} describes a compact and efficient algorithm calculating the unlinkability indicator in C-ITS and implementing previous findings to improve unlinkability. Finally, we discuss our results, their importance, novelty, advantages, and limitations in \cref{sec:disc}.

\section{\uppercase{Preliminaries}}\label{sec:prelim}

To justify subsequent modelling steps better, we introduce contextual information supporting our aim, settings and privacy assumptions.

\subsection{Aim of the study}\label{sec:aim}

To specify the aim, we analyze relevant cybersecurity requirements. Here, we use some of the classical definitions for privacy in complex systems to specify our aim with greater precision. Privacy requirements for C-ITS are often derived from ISO/IEC 15408-2. For example, (\cite{itswg5IntelligentTransportSystems2021}) suggests that the combination of pseudonymity and unlinkability offers the appropriate sender privacy protection for basic ITS safety messages (such as CAM). In simple terms, \textit{pseudonymity} requires that the identity of a user is never revealed or inferred. However, one of the major complications in dealing with pseudonymity is the following: an attacker may learn the user's identity composition based on multiple sessions, events, or traces. \textit{Unlinkability} is the assurance about the ability to resist learning such a composition (\cite{iso/iecjtc1/sc27EvaluationCriteriaIT2022}):

{\dft[Unlinkability of operations]{Requires that users and/or subjects are unable to determine whether the same user caused certain specific operations in the system, or whether operations are related in some other manner.}\label{dft:unlink_g}}

In the context of V2X communication in C-ITS with many users, the cryptographically signed messages broadcasted by the ITS-SUs (controlled by these users) should have the property of \cref{dft:unlink_g} (\cite{itswg5IntelligentTransportSystems2021,hicksVehicularDAAScheme2020}). Nonetheless, such interpretation has certain disadvantages, major of which is inflexibility. Indeed, \textit{`...unable to determine...'} statement can either be false or true, meaning that the unlinkability of the whole C-ITS (with many cars and observable during many hours) is expressed using a binary value. This issue has been recognized by practitioners and researchers alike, which is reflected in comments and best practice recommendations to ITS engineers and managers. For example, (\cite{iso/tc22/sc32CybersecurityEngineering2021}) contains the table \textit{`Example privacy impact rating criteria'}: it includes \textit{Impact rating Criteria} interpreting the meaning for the severity degrees (e.g., Negligible, Moderate, Major, Severe) for \textit{privacy impact rating} indicator. Importantly enough, interpretations in this table evolve around two aspects: a) the level of sensitivity of the information about road user; b) \textit{how easily} it can be linked to a PII (Personally Identifiable Information) principal. Such emphasis on the \textit{easiness of linking} motivates us to modify \cref{dft:unlink_g} in the following manner:

{\dft[Unlinkability of operations*]{Is the degree of inability to determine (by users and/or subjects) whether the same user caused certain specific operations in the system, or whether operations are related in some other manner.}\label{dft:unlink_g1}}

To provide convenience in comparing unlinkability in C-ITS under different conditions, we use Shannon entropy: it is an integral criterion of uncertainty in a system that fully captures the \textit{`...degree of inability to determine...'} (\cite{wagnerTechnicalPrivacyMetrics2018}).

Henceforth, the {\bf \textit{main aim}} of our paper is to develop a methodology providing high level of assurance that \textit{entropy for CAMs' origins} is high in C-ITS. 

\subsection{Settings for the study}

\begin{figure}[!h]
  \centering
   {\epsfig{file = 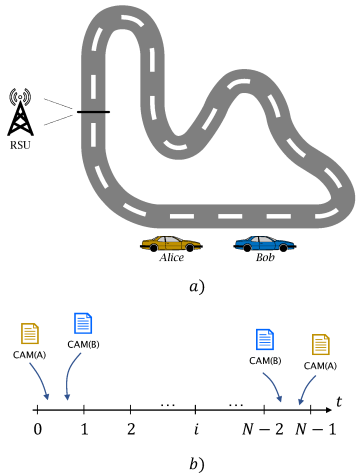, scale=0.35}}
  \caption{Setting for our study of unlinkability of CAMs.}
  \label{fig:V2X_schema}
 \end{figure}

In \cref{fig:V2X_schema}, we introduce a general setup for our study. In \cref{fig:V2X_schema}(a), two cars (ITS-SUs) are driven by \textit{Alice} and \textit{Bob}, respectively. Both ITS-SUs transmit CAMs with the same frequency, and the roadside unit (RSU) receives them without losses. The role of the \textit{attacker} is played by the RSU, who tries to separate CAMs of \textit{Alice} from CAMs of \textit{Bob}: this allows the attacker to \textit{link} CAMs belonging to the same entity. Although CAMs are signed, in our study we assume that a signature scheme providing unlinkability is used. Therefore, the separation is done based on the content of the CAMs (since the are transmitted unencrpyted) and the order of arrival of CAMs within each time interval -- see \cref{fig:V2X_schema}(b). We consider the ordering of CAMs' arrivals to be non-uniform in general: for example, in one extreme case, the CAM from \textit{Alice} arrives first, and \textit{Bob}'s CAM arrives second at any time interval $i$. If an attacker knows about such a unique property, he can link CAMs without considering their payload. However, these cases are unlikely, meaning that an attacker should also be able to infer the source (e.g., `from \textit{Alice}' or `from \textit{Bob}') of a CAM based on its content. The requirements for the content of CAMs can be found in (\cite{itswg1IntelligentTransportSystems2019}). In particular, we consider that geo-position, velocity and acceleration are essential. On the one hand, these parameters are mandatory for the \texttt{HighFrequency} container in CAM. On the other hand, numerous techniques using these parameters have been developed for the domain of Multiple-Target Tracking (MTT): corresponding estimators can be of great use for reasoning about privacy in C-ITS (\cite{blackmanMultipletargetTrackingRadar1986,karimemaraBeaconbasedVehicleTracking2013}).   

In this study, CAMs' content unlinkability is the core of our attention. We exclude from further consideration the following CAM payload: \textit{1)} cryptographically produced proofs of authenticity (e.g., signatures); \textit{2)} categorical data (e.g., vehicle role). These exclusions are due to substantial attention to issue `\textit{1)}' among the members of the cryptographic community. For example, pseudonym unlinking solutions were proposed in (\cite{camenischZoneEncryptionAnonymous2020,hicksVehicularDAAScheme2020}). Nevertheless, there is a need to complement these efforts by our study: the absence of encryption (due to safety reasons) in CAMs makes pseudonym unlinking necessary but \textit{not sufficient} for CAMs' unlinkability. This is because other data, such as geographic positions, in CAMs may be used for linking. Issue `\textit{2)}' can be omitted since categorical data is a part of \texttt{basicVehicleContainerLowFrequency} and is \texttt{OPTIONAL} in CAMs (\cite{itswg1IntelligentTransportSystems2019}). We also exclude \texttt{VehicleLength} and \texttt{VehicleWidth} (compulsory for the \texttt{HighFrequency} container), which otherwise are likely to be of great use in discriminating different vehicles (\cite{escherHowWellCan2021a}). For such an exclusion we find justification in (\cite{itswg1IntelligentTransportSystems2018}) which allows usage of the codes $1 023$ and $62$ for the length and width, respectively, if the corresponding information is unavailable.

Because of the details described above, we will model CAM as a vector in $\mathbb{R}^z$ where $z\ge 1$. Such a step is beneficial: we can apply commonly used distortion measures such as, for example, Squared Error (SE). This is a clear and straightforward way to refer to the quality degradation of essential location services (\cite{shokriPrivacyGamesLocation2016}). We, nevertheless, refrain from further discussions about the chosen distortion measure in this paper.

\subsection{Privacy assumptions and threats}

Here we provide a high-level intuition for the system and the threat of \textit{linkability}, while the details will be introduced in the subsequent sections. \textit{Alice} and \textit{Bob} coordinate their efforts. They distribute the total allowed distortion among $N-1$ time steps: as a result, they know the distortion limit for every time step $i$. At the beginning of every time interval $i$, \textit{Alice} and \textit{Bob} know the true measurements (including position, speed, acceleration, etc.) of each other. To obfuscate data in their CAMs they randomly agree on the order of their arrival at RSU at every $i$. For every $i$ they define a joint distribution according to which they change (obfuscate) their actual measurements: in expectation, they remain within the distortion limits.

An \textit{attacker} who fully controls RSU statistically infers the source of every pair of CAMs which he observes during time $i$: this statistical inference is used to calculate entropy and aligns with \cref{dft:unlink_g1}. For this, the attacker refers to the joint distribution used by \textit{Alice} and \textit{Bob} during the obfuscation. He also knows other information, such as the original geo-positions of the players at every $i$, and the probabilities for the order of CAMs' arrivals. The resulting unlinkability in the system depends on: i) statistics for the order of arrival of CAMs from the players; ii) the level of allowed distortion; iii) how far apart actual measurements of \textit{Alice} and \textit{Bob} are at every $i$.

%\subsection{Anticipated novelty}

\section{\uppercase{Mathematical model}}\label{sec:m_mod}
We explain our mathematical model in the following sections. To easy the reading, \cref{tab:varbls} contains an overview about our notations.

\begin{table}[h]\caption{Notations} \label{tab:varbls}
\centering\scalebox{0.5}{
 \begin{tabular}{|l |l |} 
 \hline
 \textbf{Notation} & \textbf{Description} \\ [0.5ex] 
 \hline\hline
 ITS & Intelligent Transport Systems\\\hline
 C-ITS & Cooperative Intelligent Transport Systems\\\hline
 ITS-SU & ITS Station Unit (including installed in vehicles)\\\hline
 V2X & Vehicle-to-Everything\\\hline
 CAM & Cooperative Awareness Message\\\hline
 RSU & Roadside Unit \\\hline
 HMM & Hidden Markov Model \\\hline
 ${\bf D}_u$& Set of user-related data\\\hline
 ${\bf D}_s$& Set of information system-related data\\\hline
 ${\bf U}$& Set of information system's users\\\hline
 ${\bf P}$& Set of data processing procedures at the information system\\\hline
 $\mathbb{P}$ & Set of players including \textit{Alice} and \textit{Bob}\\\hline
 $x^A_k,\; 1\le k\le\mu$ & A hidden state for \textit{Alice}\\\hline
 $\mathbb{X}^A =\{ x^A_k\}$ & Set of hidden states for \textit{Alice}\\\hline
 $x^B_j,\; 1\le j\le\omega$ & A hidden state for \textit{Bob}\\\hline
 $\mathbb{X}^B =\{ x^B_j\}$ & Set of hidden states for \textit{Bob}\\\hline
 $\mathbb{X}^{(A,B)}$ & Set of joint hidden states for $\bigl<Alice,\;Bob\bigr>$\\\hline
 $\mathbb{X}^{(B,A)}$ & Set of joint hidden states for $\bigl<Bob,\;Alice\bigr>$\\\hline
 $\mathcal{R}$ & Index (label) for rose nodes\\\hline
 $\mathbb{X}_{\mathcal{R}} =\mathbb{X}^{(A,B)}$ & Set of all rose nodes\\\hline
 $\mathcal{B}$ & Index (label) for blue nodes\\\hline
 $\mathbb{X}_{\mathcal{B}} =\mathbb{X}^{(B,A)}$ & Set of all blue nodes\\\hline
 $\mathbb{L} = \{\mathcal{R}, \mathcal{B}\}$ & Set of labels encoding $\left|\mathbb{P}\right|!$ combinations\\\hline
 $\mathbb{Y}$ & Set of joint observable states for \textit{Alice} and \textit{Bob}\\\hline
 $i\in\{1, 2, ..., N-1\}$ & Time-step in discrete HMM\\\hline
 $X^A_i$ & Variable on $\mathbb{X}^A$ at $i$\\\hline
 $X^B_i$ & Variable on $\mathbb{X}^B$ at $i$\\\hline
 ${\bf X}_i$ & Variable for joint hidden state on step $i$\\\hline
 $\ell_i$ & Variable on $\mathbb{L}$ at $i$\\\hline 
 ${\bf Y}_i$ & Variable on $\mathbb{Y}$ on step $i$\\\hline
 $\mathrm{Pr}\left({\bf X}_{i+1}\mid{\bf X}_i\right)$ & Probability of transition between hidden states\\\hline
 $\mathrm{Pr}\left({\bf Y}_{i}\mid{\bf X}_i\right)$ & Conditional probability for observable states\\\hline
 $\varphi$ & Order mixing (label permuting) probability\\\hline
 $\rho_i$ & Distribution over hidden states on step $i$\\\hline
 $\rho_{i+1}|\rho_i$ & Conditional distribution over hidden states on step $i+1$\\\hline
 \end{tabular}}
\end{table}

\subsection{Model setup}

We consider the generic case of information systems with the passive attacker (\emph{Eve}). An information system processes data set ${\bf D}_u$ for certain users, who form set ${\bf U}$. For this, information system executes a set of data processing algorithms ${\bf P}$: data ${\bf D}_s$, which describes configuration and parameters of these processing algorithms. Any data processing algorithm $\forall{p} \in {\bf P}$ is triggered by either a user or information system's state described by its configuration and parameters. At the data processing algorithm $p$, specific user data is taken for input that results in creation of new or altering of existing user's data as well as change of information system's configuration and parameters. In certain cases data processing algorithms within system can alter even user set ${\bf U}$. Therefore, $p$ is a mapping $p:{\bf U} \times {\bf D} \to {\bf U} \times {\bf D}$, where ${\bf D} = {\bf D}_{u} \cup {\bf D}_s$. Thus, information system is described as a tuple $\{{\bf U}, {\bf D}, {\bf P}, \emph{Eve}\}$.

For $u_\iota \in {\bf U}$ user's identification information system performs special data processing algorithms $identify_s(\cdot) \in {\bf P}$, so that $identify_s({d_u}_\iota) = u_\iota \in {\bf U}$. To protect it's data the information system may apply obfuscation algorithm $obfuscate(\cdot) \in {\bf P}$ to achieve such data alteration $d^*_{u_\iota} = obfuscate(d_{u_\iota})$ that $identify_s({d^*_u}_\iota) \neq u_\iota$.

Let's consider \emph{Eve} is watching over the data processing flow with a certain ability level that allows her to get access to the data of the information system -- $\emph{Eve} \emph{ sees } d^*_u, d^*_s$, where $d^*_u \subseteq {\bf D}_u$ and $d^*_s \subseteq {\bf D}_s$. For the unlinkability property of user's data within the considered system it is necessary that $\forall{{d^*_u}_\iota} \subseteq{\bf D}_u$ so that $\emph{Eve sees }{d^*_u}_\iota$, it is infeasible for \emph{Eve} to find such a transformation $identify_e({d_u}_\iota)$ that $identify_e({d^*_u}_\iota) = u_\iota$.

The following interpretations are possible in the context of unlinkability for C-ITS. $d_{ u } = \{registrationNumber, manufacturer, ..., PATHS\}$, where $\forall{path_\iota} \in PATHS$, $path_\iota = \{\langle x_j, y_j \rangle\}$. Some ways of user \textit{linking} are:

- $search(registrationNumber) = u_\iota$: is usually performed by law enforcement agencies;

- $recon(manufacturer, color) = u_\iota$: can be performed by private detectives, for instance;

- $drivingModelling(paths_\iota) = u_\iota$: can be performed by gathering of data from roadside units.

The research focuses on the latter way of user linking. Therefore we considering the states of users vehicles at different moments in time (communicated in CAMs) and assessing unlinkability in C-ITS through \textit{Eve}'s ability to infer information using, for instance, $drivingModelling(\cdot)$ as $identify_e(\cdot)$.

\subsection{Markov model for unlinkability}

\begin{figure}[!h]
  \centering
   {\epsfig{file = 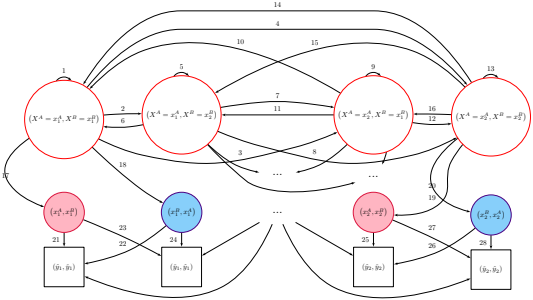, scale=0.55}}
  \caption{Hidden Markov Model for $2$ players sending CAMs.}
  \label{fig:markov}
 \end{figure}

To study unlinkability in V2X we use Hidden Markov Model which graphical representation is given on \cref{fig:markov}.
The following sets are needed to describe the model. The set of all players is $\mathbb{P} = \left\{Alice,\; Bob, ... \right\}$. For each player, there exists a set of hidden states for his vehicle, e.g., for \textit{Alice} there is $\mathbb{X}^A = \left\{ x_1^A, x_2^A, ..., x_k^A, ..., x_{\mu}^A\right\}$ and for \textit{Bob} there is $\mathbb{X}^B = \left\{ x_1^B, x_2^B, ..., x_j^B, ..., x_{\omega}^B\right\}$. Each state, for example, $x_1^A$ can be a vector including specific position, velocity, acceleration and other characteristics applicable to \textit{Alice's} vehicle at certain time. Throughout the paper we assume that $\mathbb{X}^A\cap\mathbb{X}^B$ is in general non-empty.

The system of $|\mathbb{P}|$ players is characterized by hidden and observable joint states. Transition happens between hidden states ${\bf X}_i$ and ${\bf X}_{i+1}$ when time step $i$ proceeds to $i+1$, where joint state ${\bf X}_i = \left( X_i^A, X_i^B\right)$ is the composition (concatenation) of variables $X_i^A\in\mathbb{X}^A$ and $X_i^B\in\mathbb{X}^B$ . As such, $\forall{k,j} ( x_k^A, x_j^B) \in\mathbb{X}^{(A,B)}$, where $|\mathbb{X}^{(A,B)}| = |\mathbb{X}^{A}|\times |\mathbb{X}^{B}|$ (for simplicity of representation we further assume $|\mathbb{P}|=2$, $|\mathbb{X}^{A}| =\mu =2$, $|\mathbb{X}^{B}| =\omega =2$).

Possible transitions from ${\bf X}_i$ to ${\bf X}_{i+1}$ are denoted using indices $1-16$ (see \cref{fig:markov}): these transitions are governed by corresponding probabilities. For example, the transition from ${\bf X}_i = \left( X_i^A = x_1^A, X_i^B = x_1^B\right)$ to ${\bf X}_{i+1} = \left( X_{i+1}^A = x_2^A, X_{i+1}^B = x_2^B\right)$ is denoted by index $4$. The probability of such a transition is $\mathrm{Pr}\left( X_{i+1}^A = x_2^A, X_{i+1}^B = x_2^B\mid X_i^A = x_1^A, X_i^B = x_1^B\right)$. In practice, these probabilities can be obtained based on the well-studied physical models for vehicles (\cite{blackmanMultipletargetTrackingRadar1986}).

For each ${\bf X}_i$ of the hidden joint states there are $|\mathbb{P}|!$ possible permutations for its concatenated components originating from the users. These permutations are the major cause of uncertainty when an attacker attempts to label combined CAMs of \textit{Alice} and \textit{Bob}. In practice, this is caused by the unpredictable arrangement of CAMs within each scan (or session) $i$. Hence, a permutation should be selected by randomly following one of the possible transitions. For example, while the system is in a joint state $\left( X_i^A = x_1^A, X_i^B = x_1^B\right)$ permutation $\left( x_1^A, x_1^B\right)$ (rose colored node) should be considered if transition with index $17$ takes place, and $\left(x_1^B, x_1^A\right)$ (blue coloured node) should be considered if transition $18$ happens (see \cref{fig:markov}). We will use notations ${\bf X}_{i,\mathcal{R}}$ and ${\bf X}_{i,\mathcal{B}}$ for rose and blue nodes, respectively, where ${\bf X}_{i,\mathcal{R}} \in \mathbb{X}_{\mathcal{R}}$, ${\bf X}_{i,\mathcal{B}} \in \mathbb{X}_{\mathcal{B}}$, and $\mathbb{X}_{\mathcal{R}} = \mathbb{X}^{(A,B)}$, $\mathbb{X}_{\mathcal{B}} = \mathbb{X}^{(B,A)}$. Further in the text, we will refer to the states represented by the coloured nodes as `labelled states'. For the sake of simplicity and without loss of generality, for all realizations of hidden states ${\bf X}_{i}$, we consider $\mathrm{Pr}\left({\bf X}_{i,\mathcal{R}}|{\bf X}_{i}\right)=\varphi\le 0.5$, and $\mathrm{Pr}\left({\bf X}_{i,\mathcal{B}}|{\bf X}_{i}\right)=1 -\varphi$.

To denote the totality of hidden permuted joint states we use set $\mathbb{X}_{\{\mathcal{R}, \mathcal{B}\}} = \mathbb{X}_{\mathcal{R}}\cup\mathbb{X}_{\mathcal{B}}$, where $|\mathbb{X}_{\mathcal{R}}|\le|\mathbb{X}_{\{\mathcal{R}, \mathcal{B}\}}|\le 2|\mathbb{X}_{\mathcal{R}}|$. For every ${\bf X}_{i,\mathcal{R}}$ and ${\bf X}_{i,\mathcal{B}}$ there are transitions to observable joint states ${\bf Y}_{i}\in \mathbb{Y}$, $\mathbb{Y}=\bigl\{(\hat{y}_1, \check{y}_1), (\check{y}_1, \hat{y}_1),..., (\hat{y}_q, \check{y}_q), (\check{y}_q, \hat{y}_q)$ $,...,$ $(\hat{y}_\xi, \check{y}_\xi), (\check{y}_\xi, \hat{y}_\xi) \bigr\}$. Some of these transitions to observable states are denoted with indices $21 - 28$ on \cref{fig:markov}. Until proven otherwise, the cardinality of $\mathbb{Y}$ is considered independent on $|\mathbb{X}_{\{\mathcal{R}, \mathcal{B}\}}|$.

Measuring uncertainty about label $\ell\in\mathbb{L}$, $\mathbb{L}=\{\mathcal{R}, \mathcal{B}\}$, is of our main interest: this is done based on observable states.

\section{\uppercase{Model properties}}\label{sec:mod_prop}

To formally express unlinkability following \cref{dft:unlink_g1} we will use conditional entropy $H\left( \ell_1, \ell_2,...| {\bf Y}_1, {\bf Y}_2,...\right)$ for the sequence of labels $\ell_1, \ell_2,..., \ell_{N-1}$ given that an attacker observes ${\bf Y}_1, {\bf Y}_2,..., {\bf Y}_{N-1}$ (\cite{wagnerTechnicalPrivacyMetrics2018}).

\subsection{General expression for unlinkability}

For the described HMM, probability of any hidden state at any time step can be specified using multivariate discrete distribution $\pmb{\rho}: \mathbb{X}^{(A,B)}\times\{0, 1, ..., N-1\} \rightarrow [0, 1]^{N\left|\mathbb{X}^{(A,B)}\right|}$. We will further use $\rho_i$ slices of $\pmb{\rho}$ such that $\pmb{\rho} =\bigcup\limits_{i=0}^{N-1}\rho_i$, where each slice represents a distribution over hidden states at step $i$. Slice $\rho_0$ defines distribution over the hidden states before the start of the system. Because HMM has been previously defined (see \cref{fig:markov}) using transitional probabilities that remain unchanged for all time steps, each slice can be fully determined in a conditioned sequential manner: $\rho_{i+1}|\rho_i$ means that $\rho_{i+1}$ is trivially derived if $\rho_i$ is given.

Since an attacker observes ${\bf Y}_1, {\bf Y}_2,..., {\bf Y}_{N-1}$ and knows $\pmb{\rho}$ analysis of $H\left( \ell_1, \ell_2,... \mid {\bf Y}_1, {\bf Y}_2,..., \pmb{\rho}\right)$ is central to our reasoning about unlinkability. We state the following. 

{\lmm{Unlinkability in V2X system (as per \cref{fig:markov}) is expressed as:

\begin{equation}\label{eqn:HMM_cond_01}
\begin{scalebox}{0.95}{$\!\begin{array}{lll}
H\bigl( \ell_1, \ell_2,... \big| {\bf Y}_1, {\bf Y}_2,..., \pmb{\rho}\bigr)=&\\
\sum\limits_{i=0}^{N-2}H\bigl( \ell_{i+1} \big| {\bf Y}_{i+1}, \{\rho_{i+1}| \rho_{i}\}\bigr)\;.&
\end{array}$}
\end{scalebox}
\end{equation}

}\label{lmm:unlk_entr}}

\begin{proof}
For simplicity, we consider $N=3$ only. First, it should be noted that

\begin{equation}\label{eqn:HMM_cond_02}
\begin{scalebox}{0.95}{$\!\begin{array}{lll}
H\bigl( \ell_1, \ell_2\big| {\bf Y}_1, {\bf Y}_2, \pmb{\rho}\bigr)=&\\
H\bigl( \ell_1, \ell_2, {\bf Y}_1, {\bf Y}_2 \big|\pmb{\rho}\bigr) - H\bigl({\bf Y}_1, {\bf Y}_2\big|\pmb{\rho}\bigr)\;.&
\end{array}$}
\end{scalebox}
\end{equation}

\noindent We then ponder at the right-hand side of the \cref{eqn:HMM_cond_02}. Each of these terms can be expressed as:

\begin{equation}\label{eqn:HMM_cond_03}
\begin{scalebox}{0.95}{$\!\begin{array}{lll}
H\bigl( \ell_1, \ell_2, {\bf Y}_1, {\bf Y}_2 \big|\pmb{\rho}\bigr)=&\\
H\bigl( \ell_2, {\bf Y}_2 \big| \ell_1, {\bf Y}_1, \pmb{\rho}\bigr) + H\bigl( \ell_1, {\bf Y}_1 \big| \pmb{\rho}\bigr)\;,&
\end{array}$}
\end{scalebox}
\end{equation}

\noindent and

\begin{equation}\label{eqn:HMM_cond_04}
\begin{scalebox}{0.95}{$\!\begin{array}{lll}
H\bigl({\bf Y}_1, {\bf Y}_2\big|\pmb{\rho}\bigr) = H\bigl({\bf Y}_2\big|{\bf Y}_1, \pmb{\rho}\bigr) + H\bigl({\bf Y}_1\big|\pmb{\rho}\bigr)\;,
\end{array}$}
\end{scalebox}
\end{equation}

\noindent respectively. We point out that $H\bigl( \ell_2, {\bf Y}_2 \big| \ell_1, {\bf Y}_1, \pmb{\rho}\bigr) = H\bigl( \ell_2, {\bf Y}_2 \big|\pmb{\rho}\bigr)$ and $H\bigl({\bf Y}_2\big|{\bf Y}_1, \pmb{\rho}\bigr) = H\bigl({\bf Y}_2\big|\pmb{\rho}\bigr)$ in \cref{eqn:HMM_cond_02,eqn:HMM_cond_03}, respectively. This follows from the fact that realizations of $\ell_i, {\bf Y}_i$ do not affect $\ell_{i+1}, {\bf Y}_{i+1}$. We finally stress that $\pmb{\rho}$ is redundant for determining $\ell_{i+1}, {\bf Y}_{i+1}$ since only $\rho_{i+1}|\rho_i$ has relevance: $H\bigl({\bf Y}_1\big|\pmb{\rho}\bigr) = H\bigl({\bf Y}_1\big|\{\rho_1|\rho_0\}\bigr)$, $H\bigl(\ell_1, {\bf Y}_1\big|\pmb{\rho}\bigr) = H\bigl(\ell_1, {\bf Y}_1\big|\{\rho_1|\rho_0\}\bigr)$, $H\bigl({\bf Y}_2\big|\pmb{\rho}\bigr) = H\bigl({\bf Y}_2\big|\{\rho_2|\rho_1\}\bigr)$, $H\bigl(\ell_2, {\bf Y}_2\big|\pmb{\rho}\bigr) = H\bigl(\ell_2, {\bf Y}_2\big|\{\rho_2|\rho_1\}\bigr)$. Hence, \cref{eqn:HMM_cond_02} can be rewritten as

\begin{equation}\label{eqn:HMM_cond_05}
\begin{scalebox}{0.95}{$\!\begin{array}{lll}
H\bigl( \ell_1, \ell_2\big| {\bf Y}_1, {\bf Y}_2, \pmb{\rho}\bigr)=&\\
H\bigl(\ell_1, {\bf Y}_1\big|\{\rho_1|\rho_0\}\bigr) + H\bigl(\ell_2, {\bf Y}_2\big|\{\rho_2|\rho_1\}\bigr) -&\\
\Bigl( H\bigl({\bf Y}_1\big|\{\rho_1|\rho_0\}\bigr) + H\bigl({\bf Y}_2\big|\{\rho_2|\rho_1\}\bigr) \Bigr)\;.&
\end{array}$}
\end{scalebox}
\end{equation}

The latter \cref{eqn:HMM_cond_05} can be regrouped

\begin{equation}\label{eqn:HMM_cond_06}
\begin{scalebox}{0.95}{$\!\begin{array}{lll}
H\bigl( \ell_1, \ell_2\big| {\bf Y}_1, {\bf Y}_2, \pmb{\rho}\bigr)=&\\
\Bigl(H\bigl(\ell_1, {\bf Y}_1\big|\{\rho_1|\rho_0\}\bigr) - H\bigl({\bf Y}_1\big|\{\rho_1|\rho_0\}\bigr)\Bigr)+&\\
\Bigl(H\bigl(\ell_2, {\bf Y}_2\big|\{\rho_2|\rho_1\}\bigr) - H\bigl({\bf Y}_2\big|\{\rho_2|\rho_1\}\bigr)\Bigr)=&\\
H\bigl(\ell_1 \big|{\bf Y}_1, \{\rho_1|\rho_0\}\bigr) + H\bigl(\ell_2\big|{\bf Y}_2, \{\rho_2|\rho_1\}\bigr)\;.&
\end{array}$}
\end{scalebox}
\end{equation}

\end{proof}

\subsection{Worst-case unlinkability}

We aim to obtain a computationally feasible estimation of unlinkability. Direct utilization of the results of \cref{lmm:unlk_entr} presupposes computing $\{\rho_{i+1}| \rho_{i}\}$ which has several disadvantages: \textit{a)} transition probabilities for hidden states need to be specified (which usually requires studying physical models of movement for the users); \textit{b)} total computational complexity for defining distributions over the hidden states is therefore $O(N\mu^2 \omega^2)$. To avoid these complications, we develop our unlinkability assurance based on a \textit{rational lower bound} $\mathcal{H}_{\mathrm{r}}$ for $H\bigl( \ell_1, \ell_2,... \big| {\bf Y}_1, {\bf Y}_2,..., \pmb{\rho}\bigr)$. The concept of the rational lower bound is explained through the following assumptions (\cite{sniedovichWaldMightyMaximin2016}).

{\ass[Worst-case unlinkability]{Requires that an attacker knows sets for the hidden, labelled and observable states. He knows all the transitions and the order mixing probability $\varphi$. For each observable state at time $i$ he then defines the worst possible hidden state(s) which does not contradict his knowledge.}\label{ass:lb}}

We nevertheless stress that despite \cref{ass:lb} might be viewed as excessive, the attacker does not know the labelled state $\ell_i$ (and can not force its selection) at time $i$.  

{\ass[Rational lower bound $\mathcal{H}_{\mathrm{r}}$]{Requires that users are rational and maximize worst-case unlinkability: observable states are obtained through rational obfuscation of the worst labelled states considered by the attacker.}\label{ass:rlb}}

There are several aspects affecting the task of calculating such $\mathcal{H}_{\mathrm{r}}$: 1) probabilities for transitions between hidden states (e.g., the probabilities defining $\pmb{\rho}$); 2) probabilities for transitions from the hidden states to the labelled states (e.g., $\varphi$, $1-\varphi$), and from the labelled states to the observable states. Further, we consider a situation where the worst case $\pmb{\rho}$ (\textit{minimizing entropy}) is defined for 1) while the most optimal probabilities (\textit{maximizing entropy}) are then specified for 2) under constraint $\tilde{D}$ on the total distortion over $N-1$ steps. 

We use the results of \cref{lmm:unlk_entr} to require the following:

\begin{equation}\label{eqn:WC_cond_01}
\begin{scalebox}{0.95}{$\!\begin{array}{lll}
\mathcal{H}_{\mathrm{r}} =\underset{\pmb{\rho}}{\min}\Bigl[ H\bigl( \ell_1, \ell_2,... \big| {\bf Y}_1, {\bf Y}_2,..., \pmb{\rho}\bigr)\Bigr]=&\\
\sum\limits_{i=0}^{N-2} \underset{\{\rho_{i+1}| \rho_{i}\}}{\min} \Bigl[H\bigl( \ell_{i+1} \big| {\bf Y}_{i+1}, \{\rho_{i+1}| \rho_{i}\}\bigr)\Bigr]\;.&
\end{array}$}
\end{scalebox}
\end{equation}

To obfuscate hidden states in the way maximizing $\mathcal{H}_{\mathrm{r}}$ we need to determine properties of 

\begin{equation}\label{eqn:WC_cond_02}
\scalebox{0.9}{$\rho_{\mathrm{min}, i+1}= \arg\underset{\{\rho_{i+1}| \rho_{i}\}}{\min} \Bigl[H\bigl( \ell_{i+1} \big| {\bf Y}_{i+1}, \{\rho_{i+1}| \rho_{i}\}\bigr)\Bigr]\;.$}
\end{equation}

Probabilities $\mathrm{Pr}\left( \ell_{i+1}=\mathcal{R}, {\bf Y}_{i+1}\big|\{\rho_{i+1}| \rho_{i}\}\right)$, $\mathrm{Pr}\left( \ell_{i+1}=\mathcal{B}, {\bf Y}_{i+1}\big|\{\rho_{i+1}| \rho_{i}\}\right)$ will be used in our further derivations. To simplify notations we will use $\mathrm{Pr}\left( \ell_{i+1}=\mathcal{R}, {\bf Y}_{i+1}\right)$, $\mathrm{Pr}\left( \ell_{i+1}=\mathcal{B}, {\bf Y}_{i+1}\right)$, respectively. The probabilities are defined as:

\begin{equation}\label{eqn:PR_cond_01}
\begin{scalebox}{0.9}{$\!\begin{array}{lll}
\mathrm{Pr}\left( \ell_{i+1} = \mathcal{R}, {\bf Y}_{i+1}\right) =& \\
\sum\limits_{{\bf X}_{i+1, \mathcal{R}}\in\mathbb{X}_{\mathcal{R}}}{\mathrm{Pr}\left({\bf Y}_{i+1}\mid {\bf X}_{i+1, \mathcal{R}}\right) \mathrm{Pr}\left( {\bf X}_{i+1, \mathcal{R}}\right)}=&\\
\sum\limits_{{\bf X}_{i+1, \mathcal{R}}\in\mathbb{X}_{\mathcal{R}}}{\mathrm{Pr}\left({\bf Y}_{i+1}\mid {\bf X}_{i+1, \mathcal{R}}\right) \varphi\mathrm{Pr}\left( {\bf X}_{i+1}\right)}\;,&
\end{array}$}
\end{scalebox}
\end{equation}

\begin{equation}\label{eqn:PR_cond_02}
\begin{scalebox}{0.9}{$\!\begin{array}{lll}
\mathrm{Pr}\left( \ell_{i+1} = \mathcal{B}, {\bf Y}_{i+1}\right) =& \\
\sum\limits_{{\bf X}_{i+1, \mathcal{B}}\in\mathbb{X}_{\mathcal{B}}}{\mathrm{Pr}\left({\bf Y}_{i+1}\mid {\bf X}_{i+1, \mathcal{B}}\right) \mathrm{Pr}\left( {\bf X}_{i+1, \mathcal{B}}\right)}=&\\
\sum\limits_{{\bf X}_{i+1, \mathcal{B}}\in\mathbb{X}_{\mathcal{B}}}{\mathrm{Pr}\left({\bf Y}_{i+1}\mid {\bf X}_{i+1, \mathcal{B}}\right) (1-\varphi)\mathrm{Pr}\left( {\bf X}_{i+1}\right)}\;.&
\end{array}$}
\end{scalebox}
\end{equation}

We then point out that 

\begin{equation}\label{eqn:PR_cond_03}
\mathrm{Pr}\left( \ell_{i+1}=\mathcal{R} \mid {\bf Y}_{i+1}\right) = \frac{\mathrm{Pr}\left( \ell_{i+1}=\mathcal{R}, {\bf Y}_{i+1}\right)}{\mathrm{Pr}\left( {\bf Y}_{i+1}\right)}\;,
\end{equation}

\begin{equation}\label{eqn:PR_cond_04}
\mathrm{Pr}\left( \ell_{i+1}=\mathcal{B} \mid {\bf Y}_{i+1}\right) = \frac{\mathrm{Pr}\left( \ell_{i+1}=\mathcal{B}, {\bf Y}_{i+1}\right)}{\mathrm{Pr}\left( {\bf Y}_{i+1}\right)}\;,
\end{equation}

\noindent where
\begin{equation}\label{eqn:PR_cond_05}
\scalebox{0.8}{$\mathrm{Pr}\left( {\bf Y}_{i+1}\right) = \mathrm{Pr}\left( \ell_{i+1}=\mathcal{R}, {\bf Y}_{i+1}\right)+ \mathrm{Pr}\left( \ell_{i+1}=\mathcal{B}, {\bf Y}_{i+1}\right)\;.$}
\end{equation}

The following result establishes an important property of $\rho_{\mathrm{min}, i+1}$.

{\lmm{For all $i\in[1, N-1]$ distribution $\rho_{\mathrm{min}, i}$ is degenerate.}\label{lmm:dist_dg}}

\begin{proof}
We presume that $\rho_{\mathrm{min}, i}$ is non-degenerate. For simplicity and without loss of generality we consider two-point distribution $\rho^*_{\mathrm{min}, i}: \{\mathbf{x}_1,\mathbf{x}_2\}\to [0,1]^2$. For instance, realizations $\mathbf{x}_1=(x^A_1, x^B_1)$, $\mathbf{x}_2=(x^A_2, x^B_2)$ can be used. Here $\mathrm{Pr}\left(\mathbf{X}_i = \mathbf{x}_1\right)=\xi$, and $\mathrm{Pr}\left(\mathbf{X}_i = \mathbf{x}_2\right)=1-\xi$.

Minimization of conditional entropy in \cref{eqn:WC_cond_02} is equivalent to the minimization of $p_{\ell_i,\mathrm{min}}$ where

\begin{equation}\label{eqn:p_min}
\scalebox{0.75}{$p_{\ell_i,\mathrm{min}}=\min \Bigl\{\mathrm{Pr}\left( \ell_{i}=\mathcal{R} \mid {\bf Y}_{i}\right), \mathrm{Pr}\left( \ell_{i}=\mathcal{B} \mid {\bf Y}_{i}\right)\Bigr\}\; ,$}
\end{equation}

\noindent and without loss of generality, we assume that $p_{\ell_i,\mathrm{min}}=\mathrm{Pr}\left( \ell_{i}=\mathcal{R} \mid {\bf Y}_{i}\right)$. To express $p_{\ell_i,\mathrm{min}}$ we then use \cref{eqn:PR_cond_01,eqn:PR_cond_02,eqn:PR_cond_03,eqn:PR_cond_05} with the following substitutions (simplifying expressions): $\alpha_1 = \varphi\mathrm{Pr}\left( {\bf Y}_{i}\mid {\bf X}_{{i},\mathcal{R}}=(x^A_1, x^B_1)\right)$, $\beta_1 = \varphi\mathrm{Pr}\left( {\bf Y}_{i}\mid {\bf X}_{{i},\mathcal{R}}=(x^A_1, x^B_1)\right) + (1-\varphi)\mathrm{Pr}\left( {\bf Y}_{i}\mid {\bf X}_{{i},\mathcal{B}}=(x^B_1, x^A_1)\right)$, $\alpha_2 = \varphi\mathrm{Pr}\left( {\bf Y}_{i}\mid {\bf X}_{{i},\mathcal{R}}=(x^A_2, x^B_2)\right)$, $\beta_2 = \varphi\mathrm{Pr}\left( {\bf Y}_{i}\mid {\bf X}_{{i},\mathcal{R}}=(x^A_2, x^B_2)\right) + (1-\varphi)\mathrm{Pr}\left( {\bf Y}_{i}\mid {\bf X}_{{i},\mathcal{B}}=(x^B_2, x^A_2)\right)$. The minimization task is then

\begin{equation}\label{eqn:p_min_sub}
\underset{\xi}{\min}\;p_{\ell_i,\mathrm{min}}=\underset{\xi}{\min}\frac{\xi\alpha_1 + (1- \xi)\alpha_2}{\xi\beta_1 + (1- \xi)\beta_2}\; .
\end{equation}

By analyzing $\frac{\partial}{\partial\xi} p_{\ell_i,\mathrm{min}}$, we conclude that there are no local extrema for $\xi\in (0,1)$ and, hence, minimum is obtained in one of the end points, e.g., $\xi\in\{0, 1\}$.
\end{proof}

Based on the result of \cref{lmm:dist_dg}, for every ${\bf Y}_i$ there is one and only worst-case hidden state $\tilde{\bf X}_i$ (because $\mathrm{Pr}\left( \tilde{\bf X}_{i} \mid \rho_{\mathrm{min}, i}\right) = 1$). It implies the following: 

{\cor{Design of HMM where for every state (realization) in $\mathbb{Y}$ there is one and only transition from $\mathbb{X}^{(A,B)}$ explicitly satisfies \cref{ass:lb}.}\label{cor:dist_dg_01}}

Therefore, we will further adhere to such design principle and use $\tilde{\bf X}_i$ to denote hidden states. Next, we will elaborate on: \textit{a)} what is the optimal number of different observable states ${\bf Y}_i$ for every $\tilde{\bf X}_i$? \textit{b)} how should we define optimal observable states? \textit{c)} what are the probabilities of transition (from the labelled states to the observable states)?

\subsection{Requirements for the observable states}

Here we provide our analysis from the standpoints of the system that obfuscates hidden states (e.g., the system produces observable states) on behalf of \textit{Alice} and \textit{Bob}, and hence $\tilde{\bf X}_i$ is assumed to be known. The possibilities of transitions $\tilde{\bf X}_{i,\mathcal{R}} \to {\bf Y}_i$ and $\tilde{\bf X}_{i,\mathcal{B}} \to {\bf Y}_i$ imply that a non-zero distortion $\mathbb{E}\left[D_{i}\right]$ takes place:

\begin{equation}\label{eqn:dist_02}
\mathbb{E}\left[ D_i\right] = \sum\limits_{{\bf y}^{(i)}_j\in\mathbb{Y}^{(i)}}D_{i,j}\mathrm{Pr}\left( {\bf Y}_i = {\bf y}^{(i)}_j\mid\tilde{\bf X}_i\right)\;,
\end{equation}

\noindent where

\begin{equation}\label{eqn:dist_01}
\scalebox{0.85}{$\!\begin{array}{lll}
&D_{i,j} =\mathrm{Pr}\left( \ell_i = \mathcal{R}\mid {\bf Y}_i = {\bf y}^{(i)}_j\right) d\left( \tilde{\bf X}_{i,\mathcal{R}}, {\bf y}^{(i)}_j\right) + \\
&\mathrm{Pr}\left( \ell_i = \mathcal{B}\mid {\bf Y}_i = {\bf y}^{(i)}_j\right) d\left( \tilde{\bf X}_{i,\mathcal{B}}, {\bf y}^{(i)}_j\right)\;.
\end{array}$}
\end{equation}

\noindent Here $\mathbb{Y}^{(i)}$ is the set of all observable states to which transitions exist from the realizations of $\tilde{\bf X}_{i,\mathcal{R}}$ and $\tilde{\bf X}_{i,\mathcal{B}}$ at time $i$; ${\bf y}^{(i)}_j$ is an element in $\mathbb{Y}^{(i)}$; $d\left(\cdot, \cdot\right)$ is some distortion measure (e.g., SE).

The optimization effort is two-fold: \textit{i)} how shall we obtain observable states $\mathbb{Y}^{(i)}$ in a way that $\mathcal{H}_{\mathrm{r}, i}$ is maximized under constraint $\tilde{D}_i\ge\mathbb{E}\left[ D_i\right]$? \textit{ii)} how shall we define $\tilde{D}_i$ for every time step $i$ such that $\mathcal{H}_{\mathrm{r}}$ is maximized and the total distortion constraint $\tilde{D}\ge\sum_i\mathbb{E}\left[D_i\right]$ is satisfied? We start with answering question \textit{i)}, which will assist us in answering question \textit{ii)}. 

\begin{figure}[!h]
  \centering
   {\epsfig{file = 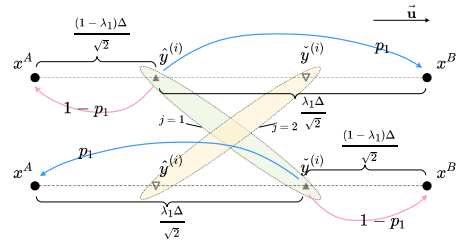, scale=0.75}}
  \caption{Scheme for obfuscation principle.}
  \label{fig:obser_st}
 \end{figure}

For the obfuscation, we utilize the following principles: every element ${\bf y}^{(i)}_j$ in $\mathbb{Y}^{(i)}$ can be fully specified by the realizations of $\tilde{\bf X}_{i,\mathcal{R}}$, $\tilde{\bf X}_{i,\mathcal{B}}$, and parameter $\lambda_j$. Probabilities $\mathrm{Pr}\left( \ell_i = \mathcal{R}\mid {\bf Y}_i = {\bf y}^{(i)}_j\right)$, $\mathrm{Pr}\left( \ell_i = \mathcal{B}\mid {\bf Y}_i = {\bf y}^{(i)}_j\right)$ then affect $\mathcal{H}_{\mathrm{r},i,j}$. All these parameters affect $D_{i,j}$. The diagram explaining relations between all the mentioned parameters is provided on \cref{fig:obser_st}. In this example, labelled states are $\tilde{\bf X}_{i,\mathcal{R}} = \left(x^A, x^B\right)$, $\tilde{\bf X}_{i,\mathcal{B}} = \left(x^B, x^A\right)$; set $\mathbb{Y}^{(i)}$ contains only two elements ${\bf y}^{(i)}_1 = \bigl( \hat{y}^{(i)}, \check{y}^{(i)}\bigr)$ and ${\bf y}^{(i)}_2 = \bigl(\check{y}^{(i)}, \hat{y}^{(i)}\bigr)$. For example, to specify ${\bf y}^{(i)}_1$ we only need $\lambda_1$ in addition to the labelled states ($\Delta$ is the distance between them). To obtain ${\bf y}^{(i)}_2$ we should apply a similar procedure where $\lambda_2$ is known (in our particular example $\lambda_2 = 1- \lambda_1$). Probability $\mathrm{Pr}\left( \ell_i = \mathcal{R}\mid {\bf Y}_i = {\bf y}^{(i)}_1\right)$ is denoted as $p_1$: its value affects attacker's uncertainty $\mathcal{H}_{\mathrm{r}, i, j}$ as well as the distortion $D_{i,j}$.

To maximize $\mathcal{H}_{\mathrm{r}, i}$ under $\tilde{D}_i\ge\mathbb{E}\left[D_i\right]$ we consider realizations of ${\bf Y}_i$ and optimal adjustment of $\lambda$: such adjustment then allows us to increase $p_1$ and $1 -p_2$.

We note that ${\bf Y}_i$ shall belong to a line segment (in a multidimensional space) connecting $\tilde{\bf X}_{i,\mathcal{R}}$ and $\tilde{\bf X}_{i,\mathcal{B}}$. This property is trivial (goes without proof) and can be best understood if triangle $\vartriangle \tilde{\bf X}_{i,\mathcal{R}}{\bf Y}_i\tilde{\bf X}_{i,\mathcal{B}}$ is considered. As a result:

\begin{equation}\label{eqn:lmbd_01}
\scalebox{0.7}{$\!\forall{{\bf Y}_i}\Bigl({\bf Y}_i\in\mathbb{Y}^{(i)}\implies \bigl(\exists\lambda\in[0, 1]\bigr)\land\bigl( \vec{{\bf Y}_i} = \overrightarrow{\tilde{\bf X}_{i,\mathcal{R}}} + \lambda\overrightarrow{\tilde{\bf X}_{i,\mathcal{R}} \tilde{\bf X}_{i,\mathcal{B}}} \bigr) \Bigr)\;.$}
\end{equation}

We then establish the following:
{\lmm{To minimize $D_{i,j}$ it is required that $\lambda_j = 1-\mathrm{Pr}\left(\ell_i =\mathcal{R}\mid {\bf Y}_i = {\bf y}^{(i)}_j\right)$.}\label{lmm:lmbd}}
\begin{proof}
From \cref{eqn:dist_01} and \cref{eqn:lmbd_01} we derive that
$$D_{i,j} = p_j\Delta^2_i\lambda^2_j + (1-p_j)\Delta^2_i(1-\lambda_j)^2\; ,$$
\noindent where $p_j=\mathrm{Pr}\left(\ell_i = \mathcal{R}\mid {\bf Y}_i={\bf y}^{(i)}_j\right)\le 0.5$, and $\Delta^2_i = d\left(\tilde{\bf X}_{i,\mathcal{R}}, \tilde{\bf X}_{i,\mathcal{B}}\right) =\left\| \overrightarrow{\tilde{\bf X}_{i,\mathcal{R}} \tilde{\bf X}_{i,\mathcal{B}}}\right\|^2$. We next analyse $\frac{\partial}{\partial \lambda_j}D_{i,j}$ and find that $\lambda_j = 1-p_j$ is the extremum (minimum) of $D_{i,j}$. 
\end{proof}

{\cor{Minimal distortion is $D_{i,j}=\Delta^2_i p_j(1-p_j)\le\frac{\Delta^2_i}{4}$, where $p_j=\mathrm{Pr}\left(\ell_i =\mathcal{R}\mid{\bf Y}_i={\bf y}^{(i)}_j\right)$, $\Delta^2_i=d\left(\tilde{\bf X}_{i,\mathcal{R}}, \tilde{\bf X}_{i,\mathcal{B}}\right)$.}\label{cor:lmbd_01}}

{\cor{For every $i$, the highest lower bound (maxmin entropy) is:
\begin{equation}\label{eqn:maxmin_entr}
\mathcal{H}_{\mathrm{r},i}= -\nu_i\log_2{\nu_i}-(1-\nu_i)\log_2{(1-\nu_i)}\; ,
\end{equation}

\noindent where $\nu_i = \min \left\{\varphi, \frac{\Delta_i - \sqrt{\Delta^2_i -4\mathbb{E}\left[ D_i\right]}}{2\Delta_i} \right\}\; .$

}\label{cor:lmbd_02}}

\begin{proof}
It is required: 1) to determine $\mathbb{Y}^{(i)}$ and probability distribution over it; 2) to determine $\mathrm{Pr}\left(\ell_i\mid{\bf Y}_i\right)$ for every element in $\mathbb{Y}^{(i)}$. For this, we demonstrate that maximum entropy under distortion constraint on $\mathbb{E}\left[ D_i\right]$ is achieved for $\left|\mathbb{Y}^{(i)} \right|\le 2$: we analyze the case for $\mathbb{Y}^{(i)}=\left\{{\bf y}_1^{(i)}, {\bf y}_2^{(i)}\right\}$ where

\begin{equation}\label{eqn:cond_03}
\mathrm{Pr}\left( \ell_i = \mathcal{R}\mid{\bf Y}_i = {\bf y}_1^{(i)}\right) = \mathrm{Pr}\left( \ell_i = \mathcal{B}\mid{\bf Y}_i = {\bf y}_2^{(i)}\right)\;.
\end{equation}

To prove the optimality of such settings, we consider several alternative cases where $\mathbb{E}\left[ D_i\right] =\tilde{D}_i$ is fixed. Let us first consider an alternative case where $\left|\mathbb{Y}^{(i)}\right| = 2$ but

\begin{equation}\label{eqn:cond_ass_01}
\resizebox{\linewidth}{!}{$\!\begin{array}{lll}
&\mathrm{Pr}\left( \ell_i = \mathcal{R}\mid{\bf Y}_i = {\bf y}_1^{(i)}\right) &\neq \mathrm{Pr}\left( \ell_i = \mathcal{R}\mid{\bf Y}_i = {\bf y}_2^{(i)}\right)\;;\\
&\mathrm{Pr}\left( \ell_i = \mathcal{R}\mid{\bf Y}_i = {\bf y}_1^{(i)}\right) &\neq \mathrm{Pr}\left( \ell_i = \mathcal{B}\mid{\bf Y}_i = {\bf y}_2^{(i)}\right)\;.
\end{array}$}
\end{equation}

For simplicity, we use the following notations: $\mathrm{Pr}\left( {\bf Y}_i = {\bf y}^{(i)}_1\mid \tilde{\bf X}_i\right) = \alpha$, and $\mathrm{Pr}\left( {\bf Y}_i = {\bf y}^{(i)}_2\mid \tilde{\bf X}_i\right) = 1-\alpha$; $\mathrm{Pr}\left( \ell_i = \mathcal{R}\mid{\bf Y}_i = {\bf y}_1^{(i)}\right) = p_1\le 0.5$, and $\mathrm{Pr}\left( \ell_i = \mathcal{R}\mid{\bf Y}_i = {\bf y}_2^{(i)}\right) = p_2\ge 0.5$. Taking into account the expression for conditional entropy, we then require:

\begin{equation}\label{eqn:req_sys_01}
\scalebox{0.9}{$\begin{cases}
\max\bigl[\mathcal{H}_{\mathrm{r},i}\bigr] &=\max\bigl[\alpha H_1 + (1-\alpha)H_2\bigr]\; ;\\
\tilde{D}_i = \alpha D_{i,1} + &(1-\alpha) D_{i,2}\;,
\end{cases}$}
\end{equation}

\noindent where $H_1=H\left(\ell_i\mid{\bf Y}_i = {\bf y}^{(i)}_1\right)$, $H_2=H\left(\ell_i\mid{\bf Y}_i = {\bf y}^{(i)}_2\right)$. Based on \cref{eqn:cond_ass_01} $D_{i,1}\neq D_{i,2}$. We now show that $H_1$ and $H_2$ are functions of $D_{i,1}$ and $D_{i,2}$, respectively. For this, we only point out that $p_1$ (similar results can be obtained for $p_2$) is a monotonically increasing function of $D_{i,1}$: it follows from \cref{cor:lmbd_01} that $p_1 = \frac{\Delta_i - \sqrt{\Delta^2_i -4D_{i,1}}}{2\Delta_i}$. To demonstrate the fallacy of attaining both \cref{eqn:cond_ass_01} and \cref{eqn:req_sys_01} it is sufficient to show the following (concavity):

\begin{equation}\label{eqn:cond_04}
\alpha F(x) + (1-\alpha) F\left(\frac{\tilde{D}_i - \alpha x}{1-\alpha}\right)\le F(\tilde{D}_i)\;,
\end{equation}

\noindent where $x=D_{i,1}$, and $F(x) = -p_1(x)\log\bigl(p_1(x)\bigr) -\bigl( 1 - p_1(x)\bigr)\log\bigl( 1 - p_1(x)\bigr)$. The validity of \cref{eqn:cond_04} follows from

\begin{equation*}
\resizebox{\linewidth}{!}{$\!\begin{array}{lll}
&\frac{\partial}{\partial x}F(x) = &\frac{1}{\Delta_i\theta}\log\left(\frac{\Delta_i +\theta}{\Delta_i -\theta}\right)\ge 0\;;\\
&\frac{\partial^2}{\partial x^2}F(x) = -\frac{2}{\Delta_i\theta^2}&\biggl(\frac{1}{\Delta_i +\theta} + \frac{1}{\Delta_i -\theta} - \frac{1}{\theta}\log\left(\frac{\Delta_i +\theta}{\Delta_i -\theta} \right) \biggr)\le 0\;,
\end{array}$}
\end{equation*}

\noindent where $\theta =\sqrt{\Delta^2_i - 4x}$, and $x\in \left[ 0, \frac{\Delta^2_i}{4}\right]$.

Next, we point out a different case where $\left|\mathbb{Y}^{(i)} \right| > 2$ and demonstrate that it is non-optimal. For this we consider $\left|\mathbb{Y}^{(i)} \right| = 3$ while the conclusions for $\left|\mathbb{Y}^{(i)} \right| > 3$ can be derived inductively then. Similarly to \cref{eqn:req_sys_01} we demand

\begin{equation*}
\scalebox{0.8}{$\begin{cases}
\max\bigl[\mathcal{H}_{\mathrm{r},i}\bigr] =\max\bigl[\alpha H_1 +&\beta H_2 + (1-\alpha -\beta) H_3\bigr]\; ;\\
\tilde{D}_i = \alpha D_{i,1} + \beta D_{i,2} + &(1-\alpha -\beta) D_{i,3}\;.
\end{cases}$}
\end{equation*}

The task is then to show that there is ${\bf y}^{(i)}_4$ for which $D_{i,4}=\frac{\alpha D_{i,1} + \beta D_{i,2}}{\alpha + \beta}$, and $\max H_4\ge \max\left[\frac{\alpha}{\alpha + \beta}H_1 + \frac{\beta}{\alpha + \beta}H_2 \right]$. We henceforth maintain that $\left|\mathbb{Y}^{(i)} \right| \le 2$ represents optimal settings.

To obtain $\max\bigl[\alpha H_1 + (1-\alpha)H_2\bigr]$ in \cref{eqn:req_sys_01} it is sufficient that $H_1 = H_2$ and $D_{i,1} = D_{i,2} =\tilde{D}_i$. The latter requires that either $\lambda_1 = \lambda_2$ or $\lambda_1 = (1-\lambda_2)$: the first condition implies $p_1 = p_2 =0.5$ and leads to a trivial situation where ${\bf y}^{(i)}_1 = {\bf y}^{(i)}_2 = 0.5\left(\tilde{\bf X}_{i, \mathcal{R}} + \tilde{\bf X}_{i, \mathcal{B}}\right)$ meaning that $\left|\mathbb{Y}^{(i)} \right|= 1$. The second condition implies $p_1 = 1- p_2$ and leads to ${\bf y}^{(i)}_1 \neq {\bf y}^{(i)}_2$ if $\tilde{D}_i < 0.25\Delta^2_i$.

Requirement $\alpha\in [0, 1]$ must be consistent with the order mixing probability $\varphi$:

\begin{equation}\label{eqn:cond_alpha}
\alpha p_1 + (1-\alpha) p_2 = \varphi\;, 
\end{equation}

\noindent from which we derive $\alpha =\frac{\varphi + p_1 -1}{2p_1 -1}$ demanding $\varphi\ge p_1$. Alternatively, this demand can be understood based on the fact $H\left(\ell_i \right)\ge H\left(\ell_i\mid{\bf Y}_i\right)$: setting $p_1 >\varphi$ results in a greater distortion, but this does not increase entropy.

\end{proof}

There are several important takeaways from the proof of \cref{cor:lmbd_02}. First, for every hidden state $\tilde{\bf X}_i$ there are two observable states that are obtained according to \cref{eqn:lmbd_01} where $\lambda^{(i)}_1 =1-\nu_i$ is used to define realisation ${\bf y}^{(i)}_1$, and $\lambda^{(i)}_2 =1-\lambda^{(i)}_1$ is used for ${\bf y}^{(i)}_2$. Second, maximum allowed distortion should be used at step $i$ meaning that $\mathbb{E}\left[D_i\right]=\tilde{D}_i$. Third, probabilities for transitions from labelled states to observable states are

\begin{equation}\label{eqn:cor_concl}
\scalebox{0.75}{$\!\begin{array}{lll}
&\mathrm{Pr}\left( {\bf Y}_i = {\bf y}^{(i)}_1 \mid \ell_i =\mathcal{R}\right) = &\frac{\nu_i}{\varphi}\frac{\varphi + \nu_i -1}{2\nu_i -1}\;;\\
&\mathrm{Pr}\left( {\bf Y}_i = {\bf y}^{(i)}_2 \mid \ell_i =\mathcal{R}\right) = &1- \mathrm{Pr}\left( {\bf Y}_i = {\bf y}^{(i)}_1 \mid \ell_i =\mathcal{R}\right)\;;\\
&\mathrm{Pr}\left( {\bf Y}_i = {\bf y}^{(i)}_1 \mid \ell_i =\mathcal{B}\right) = &\frac{1-\nu_i}{1-\varphi}\frac{\varphi + \nu_i -1}{2\nu_i -1}\;;\\
&\mathrm{Pr}\left( {\bf Y}_i = {\bf y}^{(i)}_2 \mid \ell_i =\mathcal{B}\right)= &1-\mathrm{Pr}\left( {\bf Y}_i = {\bf y}^{(i)}_1 \mid \ell_i =\mathcal{B}\right)\;.
\end{array}$}
\end{equation}

\subsection{Optimal obfuscation for $N-1$ time steps}

For every $i$ we now define $\tilde{D}_i$ such that $\mathcal{H}_{\mathrm{r}}=\sum_i \mathcal{H}_{\mathrm{r}, i}$ is maximized under the total distortion constraint $\tilde{D}\ge\sum_i\tilde{D}_i$. For this reason, we obtain optimal observable states and corresponding transition probabilities (from the labelled states) for all the time steps. From the proof of the \cref{cor:lmbd_02} we use that $\frac{\partial}{\partial {\tilde{D}_i}}\mathcal{H}_{\mathrm{r},i}\ge 0$ and $\frac{\partial^2}{\partial {\tilde{D}_i}^2}\mathcal{H}_{\mathrm{r},i}\le 0$. To maximize $\mathcal{H}_{\mathrm{r}}$ we therefore require

\begin{equation}\label{eqn:opt_req}
\scalebox{0.9}{$\begin{cases}
\forall i \;\frac{\partial}{\partial {\tilde{D}_i}}\mathcal{H}_{\mathrm{r},i} = \frac{1}{\Delta_i^2\sqrt{1-\kappa_i}}\log\left(\frac{1 +\sqrt{1-\kappa_i}}{1 -\sqrt{1-\kappa_i}}\right) =C \; ;\\
\tilde{D} = \sum\limits_{i=1}^{N-1}\tilde{D}_i = \frac{1}{4}\sum\limits_{i=1}^{N-1}\kappa_i\Delta_i^2\;,
\end{cases}$}
\end{equation}

\noindent where $C$ is some constant, $\kappa_i=\frac{4\tilde{D}_i}{\Delta_i^2}$. We then solve the system \cref{eqn:opt_req} for all $\kappa_i$, $i\in [1, N-1]$, and according to \cref{cor:lmbd_02} obtain $\nu_i = \min \left\{\varphi,\; 0.5 -\sqrt{0.25 - 0.25\kappa_i}\right\}$.

\section{\uppercase{Obfuscation algorithm}}\label{sec:obf_alg}

Here we represent our aforementioned findings in the form of obfuscation algorithm (see \cref{algo_obfsc}). It is practical and can be implemented in real settings: its complexity (excluding the complexity of \texttt{solve} procedure) is only $O(N)$. For input, the algorithm accepts arrays (of size $N$) ${\bf X}^A$, ${\bf X}^B$, and scalars $\tilde{D}$, $\varphi$. Elements of these arrays are scalar/vector realizations for $X^A_i$ and $X^B_i$ characterizing geo-positions of \textit{Alice} and \textit{Bob}, respectively, at time $i$. In practice, these arrays may contain extrapolations based on historical data and repetitive patterns. For example, \textit{Alice} and \textit{Bob} may commute to work using the same routes and roughly at the same time every day. Procedure \texttt{solve} provides a solution to \cref{eqn:opt_req}: array $\pmb{\kappa}$ contains elements $\kappa_i$ needed to define realizations for obfuscated state ${\bf Y}_i$. It is also needed to calculate the unlinkability criterion (entropy) $\mathcal{H}_{\mathrm{r},i}$ dependent on the obfuscation process. Procedure \texttt{send\_RSU} encapsulates data obfuscated at time $i$ in accordance with one of the V2X communication formats and sends it to the nearest RSU. The output of the algorithm is, therefore, an array ${\bf Y}$ containing all the obfuscated records and the indicator of the total unlinkability in the system over $N-1$ steps, $\mathcal{H}_{\mathrm{r}}$.

\IncMargin{1em}
\begin{algorithm}\scriptsize
\SetKwInOut{Input}{input}\SetKwInOut{Output}{output}
\Input{${\bf X}^A$, ${\bf X}^B$, $\tilde{D}$, $\varphi\;;$}
\Output{${\bf Y}$, $\mathcal{H}_{\mathrm{r}}\;;$}
\BlankLine
\Begin{
$\mathcal{H}_{\mathrm{r}}\leftarrow 0$, ${\bf Y}\leftarrow\varnothing$, $\pmb{\kappa}\leftarrow\mathtt{solve}\left(\tilde{D},{\bf X}^A, {\bf X}^B\right)\;;$  

\For{$i\leftarrow 1$ \KwTo $N-1$}{
$\nu_i\leftarrow\min\left\{\varphi,\; 0.5 -\sqrt{0.25 -0.25\kappa_i}\right\}$, $\alpha \leftarrow(\varphi + \nu_i -1)/(2\nu_i -1)$, $\mathcal{H}_{\mathrm{r},i}\leftarrow -\nu_i\log(\nu_i) - (1-\nu_i)\log(1-\nu_i)$, $\mathcal{H}_\mathrm{r}\leftarrow\mathcal{H}_\mathrm{r} + \mathcal{H}_{\mathrm{r},i}$, $P_{1,\mathcal{R}}\leftarrow \nu_i\alpha/\varphi$, $P_{1,\mathcal{B}}\leftarrow (1-\nu_i)\alpha/(1-\varphi)$, $r_1\leftarrow \mathtt{UniRand}\bigl( [0,1]\bigr)$, $r_2\leftarrow \mathtt{UniRand}\bigl( [0,1]\bigr)$, $\Lambda_{\mathcal{R}}\leftarrow 0.5 +(0.5 -\nu_i)\;\mathtt{sign}_{\scriptscriptstyle\pm}\!\!\left( P_{1,\mathcal{R}} - r_2\right)$, $\Lambda_{\mathcal{B}}\leftarrow 0.5 +(0.5 -\nu_i)\;\mathtt{sign}_{\scriptscriptstyle\pm}\!\!\left( P_{1,\mathcal{B}} - r_2\right)\;;$

\lIf{$r_1\le\varphi$}{$\hat{y}^{(i)}\leftarrow X^A_i + \Lambda_{\mathcal{R}}\left( X^B_i - X^A_i\right)$, $\check{y}^{(i)}\leftarrow X^B_i + \Lambda_{\mathcal{R}}\left( X^A_i - X^B_i\right)$, ${\bf Y}_i =\mathtt{concat}(\hat{y}^{(i)}, \check{y}^{(i)})$}
\lElse{$\hat{y}^{(i)}\leftarrow X^A_i + \Lambda_{\mathcal{B}}\left( X^B_i - X^A_i\right)$, $\check{y}^{(i)}\leftarrow X^B_i + \Lambda_{\mathcal{B}}\left( X^A_i - X^B_i\right)$, ${\bf Y}_i =\mathtt{concat}(\check{y}^{(i)},\hat{y}^{(i)})$}
$\mathtt{send\_RSU}\left({\bf Y}_i\right)$, ${\bf Y}=\mathtt{concat}({\bf Y}, {\bf Y}_i)\;;$
}
}
\caption{Obfuscation algorithm}\label{algo_obfsc}
\end{algorithm}\DecMargin{1em}

\section{\uppercase{Discussion}}\label{sec:disc}

Here we discuss how well the {\bf \textit{main aim}} (see \cref{sec:aim}) -- “to develop a methodology providing a high level of assurance that \textit{entropy for CAMs' origins} is high in C-ITS” -- was achieved by our paper. For this, we provide characteristics about the major results, their advantages, limitations, and plans for further work. 

\subsection{Results and their characteristics}

In this paper, we combine: \textit{(i)} the classical definition of unlinkability and \textit{(ii)} assumptions about a strong attacker to \textit{(iii)} measure and improve unlinkability in C-ITS by developing the optimal joint obfuscation technique. The \textit{academic novelty} is due to the combination of points \textit{(i-iii)}. Next, we discuss the importance of each point in greater detail. 

First, we lean towards the classical definition of \textit{unlinkability} demanded by the standards governing the domain of C-ITS applications (\cite{itswg5IntelligentTransportSystems2021,iso/tc204ITSStationManagement2018}). The concept of this study is, therefore, closer to some early works on location privacy, such as (\cite{shokriQuantifyingProtectingLocation2012}) relying on Bayesian inference in HMM and contrasts with many later works reliant on $k$-anonymity, differential privacy, and geo-indistinguishability (\cite{corserEvaluatingLocationPrivacy2016,andresGeoindistinguishabilityDifferentialPrivacy2013,bordenabeOptimalGeoIndistinguishableMechanisms2014}).

In this work, the definition of unlinkability is captured through Bayesian inference and is further reflected by entropy. We also stress on advantages of such an approach. Based on \cref{dft:unlink_g1}, the attacker's \textit{reasoning} about the operations plays the central role. Such reasoning may go beyond properties observable within the object (e.g., cryptographically signed CAM message): it may additionally rely on meta-information collected, for example, on a system level (e.g., the order of CAMs arrivals at RSU). This feature accords with many concepts in science and philosophy. Among others, Leibniz stated that indiscernible objects have identities (\cite{hackingIdentityIndiscernibles1975}): preferences about these identities can be expressed statistically (and, hence, used in Bayesian inference). For instance, in statistics such reasoning may be assisted by means of additional indexing. In contrast, geo-indistinguishability does not support further reasoning about indiscernible pieces of geo-data, making this privacy concept less demanding (and, hence, inferior) compared to unlinkability. To witness the differences between these concepts, one should observe \textit{order mixing} (and corresponding probability $\varphi$) in our HMM (see \cref{fig:markov}): even if $X_i^A = X_i^B$, the model sets $(X_i^A, X_i^B)$ apart from $(X_i^B, X_i^A)$. For that reason, we agree with the authors of (\cite{montazeriAchievingPerfectLocation2017,takbiriLimitsLocationPrivacy2017,takbiriPrivacyDependentUsers2020}), stating that both obfuscation and permutation (order mixing) are required for strong privacy assurance.

Second, our focus on unlinkability (and not on the location protection) provides consistency even under the assumption that an \textit{adversary is strong}: the attacker knows the actual locations of \textit{Alice} and \textit{Bob} at every moment $i$. He also knows the probabilistic obfuscation and order mixing algorithm used by the players. However, he does not know the outputs of this probabilistic algorithm. His goal is then to infer the origin of the digitally signed obfuscated messages. As a result of applying \cref{ass:lb}, reasoning about statistical inference made by the attacker is very much simplified compared to (\cite{shokriQuantifyingProtectingLocation2012}): information about HMM's hidden states (e.g., actual locations, velocities, etc.) and transition probabilities are not required for such reasoning. The latter detail is beneficial for the privacy assurance since establishing probabilities for transitions in HMM is a laborious and often imprecise procedure relying on Kalman-like estimators (\cite{blackmanMultipletargetTrackingRadar1986,lehmannSuboptimalKalmanFiltering2020}).

Third, strong (and simplifying) assumptions assist us in \textit{specifying} the lower bound of unlinkability in C-ITS. For every time step $i$ unlinkability is expressed through entropy $\mathcal{H}_{\mathrm{r},i}$: the worst-case inference is made by an attacker meaning that $\mathcal{H}_{\mathrm{r},i}$ is the lower bound (see \cref{ass:lb}). Components $\mathcal{H}_{\mathrm{r},i}$ are then summed over $N-1$ steps to obtain $\mathcal{H}_{\mathrm{r}}$ (see \cref{lmm:unlk_entr}). Such summation is a simple and intuitive step. It is, nevertheless, justified because for any $i\in\{1, 2, ..., N-1\}$, inference about the source (origin) of arrived CAM is independent from such inference at $i-1$. An analogy can be established between entropy (unlinkability) values $\mathcal{H}_{\mathrm{r},i}$ and $\mathcal{H}_{\mathrm{r}}$, and the concepts of microscopic and macroscopic privacy, respectively, in (\cite{shokriUnifiedFrameworkLocation2010}). Better protection of macroscopic privacy (e.g., trajectories) requires higher uncertainty about labels $\ell_i$ for the locations reported on the microscopic level (e.g., geographic points). Higher uncertainty about $\ell_i$ can only be provided at the cost of higher expected distortion $\mathbb{E}[D_i]$ of the players' CAMs. To maximize uncertainty $\mathcal{H}_{\mathrm{r},i}$ about labels under the constraint on distortions we propose a new simple way to define a \textit{joint distribution} that must be followed during the obfuscation (to obtain CAMs) conducted cooperatively by \textit{Alice} and \textit{Bob}. Optimality of (multivariate) noise parameters under various constraints on distortions has been studied by many authors in the past (\cite{andresGeoindistinguishabilityDifferentialPrivacy2013,gengOptimalNoiseAdding2016,takbiriLimitsLocationPrivacy2017}). Our approach, however, differs: to improve microscopic privacy at $i$ we insist on \textit{joint probabilistic obfuscation} of the samples from \textit{different} players (see proof of \cref{cor:lmbd_02}). Because of the latter feature, our obfuscation approach is clearly \textit{data-dependent} (\cite{croftDifferentiallyPrivateObfuscation2019,guanSuperstringBasedSequenceObfuscation2022}). We then analyse how to optimally distribute distortions over $N-1$ time steps if the corresponding distortion cap is specified for the whole duration of C-ITS observation (see \cref{eqn:opt_req}). All the findings of this paper are incorporated in \cref{algo_obfsc}. Procedure \texttt{solve} is one of the major factors contributing to the time complexity of the algorithm. This, nevertheless, can be addressed if the obfuscation optimality is slightly sacrificed. For example, \texttt{solve} can be pre-computed for several cases only: each case would produce a distinct kind of distribution for a random variable $\frac{\Delta^2_i}{\tilde{D}}$. Then, the actual input data should be approximated by the best-matching distribution, and the corresponding pre-computed outputs of \texttt{solve} should be used for the obfuscation. Such workaround can also turn our algorithm into a `real-time algorithm': if \textit{Alice} and \textit{Bob} believe that their future data will align well with one of the pre-computed distributions (e.g., because of habitual daily commutes) they can obfuscate it `on-the-fly'. Hence, the pre-computed cases for \texttt{solve} can be treated as \textit{profiles} that pairs of players agree to use.

\subsection{Limitations and future work}

This paper has certain limitations which we plan to address in our further studies.

Only one particular composition of entropy (to measure unlinkability) and squared error (to measure distortion) is considered in our work. Besides entropy, other uncertainty measures may be useful for expressing unlinkability in C-ITS (\cite{wagnerTechnicalPrivacyMetrics2018}). Also, distortion measures other than SE have been analysed and recommended in the past by some authors studying Multiple-Target Tracking and its applications (\cite{gorjiPerformanceMeasuresMultiple2011}).

Only high assurance about minimally achievable unlinkability (e.g., rational lower bound, see \cref{ass:lb,ass:rlb}) is studied here. However, to further improve the practicality of our methodology, indicators obtained for the worst-case scenario (e.g., a very strong attacker) may be complemented by other indicators relying on less pessimistic scenarios (e.g., a weaker attacker).

Only $2$ players are considered in our model. This substantially limits the number of permutations for CAMs: as a result, for every hidden state there are only $2$ labelled states (see \cref{fig:markov}). Because of that, the methodology defining observable states is also simple (see \cref{fig:obser_st}). We plan to increase the number of players in the future. However, for larger numbers of players, defining optimal observable states and corresponding probabilities for transitions is a non-trivial task. This is because more complex structures and transformations in $\mathbb{R}^z,\; z\ge 1$, need to be analyzed to optimize instant and joint obfuscation of CAMs.

\section{\uppercase{Acknowledgment}}

This research is co-financed by public funding of the state of Saxony, Germany.  
  
{\tiny \printbibliography}

\end{document}